\documentclass[aps,prb,twocolumn,showpacs,superscriptaddress]{revtex4-1}
\usepackage{amssymb}
\usepackage{amsmath}
\usepackage{upgreek}
\usepackage{graphicx}
\usepackage{hyperref}

\begin{document}

\title{Color ice states, weathervane modes, and order by disorder in the bilinear-biquadratic pyrochlore Heisenberg antiferromagnet}

\author{Yuan Wan}
\affiliation{Perimeter Institute for Theoretical Physics, Waterloo, Ontario N2L 5G7, Canada}
\author{Michel J. P. Gingras}
\affiliation{Perimeter Institute for Theoretical Physics, Waterloo, Ontario N2L 5G7, Canada}
\affiliation{Department of Physics and Astronomy, University of Waterloo, Waterloo, ON, N2L 3G1, Canada}
\affiliation{Canadian Institute for Advanced Research, 180 Dundas St. W., Toronto, ON, M5G 1Z8, Canada}

\date{\today}


\begin{abstract}
We study the pyrochlore Heisenberg antiferromagnet with additional positive biquadratic interaction in the semiclassical limit. The classical ground state manifold of the model contains an extensively large family of non-coplanar spin states known as ``color ice states''. Starting from a color ice state, a subset of spins may rotate collectively at no energy cost. Such excitation may be viewed in this three-dimensional system as a ``membrane-like'' analog of the well-known weathervane modes in the classical kagome Heisenberg antiferromagnet. We investigate the weathervane modes in detail and elucidate their physical properties. Furthermore, we study the order by disorder phenomenon in this model, focusing on the role of harmonic fluctuations. Our computationally-limited phase space search suggests that quantum fluctuations select three different states as the magnitude of the biquadratic interaction increases relative to the bilinear interaction, implying a sequence of phase transitions solely driven by fluctuations.
\end{abstract}

\maketitle


\section{Introduction}\label{sec:intro}

In simple magnetic systems with well-established classical long-range order, thermal and quantum fluctuations lead to a reduction of order parameter and are thus detrimental to the order. On the other hand, in frustrated magnetic systems~\cite{Springer_book}, where the lattice geometry and/or competing spin-spin interactions give rise to an accidentally degenerate classical ground state manifold, fluctuations can, instead, \emph{cause} long-range order by stablizing a unique ground state. While the energy of every classical ground states is the same, the excitation spectrum governing the level of thermal or quantum fluctuations about each ground state is generally different. In the classical limit, the thermal entropy of each state is different. Likewise, at zero temperature, different classical ground states receive different zero point energy corrections.  This is the engine behind the order by thermal (quantum) disorder mechanism in which a state with maximal entropy (minimal energy) gets selected out of the degenerate manifold~\cite{Villain1980,Shender1982,Kawamura1984,Henley1989,Chandra1990}. 

Given the decisive role that fluctuations may play in frustrated magnets, one may natrually ask: can there be fluctuation-driven phase transitions among classically degenerate states? It is conceivable that, as one tunes the model parameters of a frustrated magnet, the classical degenerate ground state manifold remains the same, but the thermal or quantum fluctuations select different members of the ground state manifold, thereby giving rise to a phase transition solely driven by fluctuations.

Even though order by disorder seems to be a natural setting for fluctuation-driven phase transitions, known examples appear to be relatively rare. To set the stage for the present work, it is useful to review a few related contexts. In some quantum magnetic systems such as the $XY$ pyrochlore magnet~\cite{Wong2013} and the face-centered-cubic (FCC) antiferromagnet with additional pseudo-dipolar interaction~\cite{Ishizuka2015b}, the zero-point energy correction gives rise to a quantum phase transition between different ground states as one tunes a model parameter. In both examples, the dimension of the classical ground state manifold is finite. Perhaps more unusual is the quantum phase transition in the $\mathrm{XXZ}$ kagome antiferromagnet~\cite{Chernyshev2014,Chernyshev2015}, which occurs within an extensively degenerate ground state manifold. 

There are also frustrated spin models in which the true ground state competes with an energetically metastable state. In a number of classical models~\cite{Pinettes2002,Chern2008,McClarty2014}, thermal fluctuations modify the free energy landscape and  turn the latter into a global free energy minimum over a certain temperature range. As temperature decreases, the system should, in principle, exhibit a thermal transition from the energetically metastable state to the true ground state~\footnote{One generally expects the metastable and minimum energy states to be of different symmetry. In this case, a thermodynamic transition between the two should be first order. Whether droplets of the minimum energy state can, in specific cases, be nucleated out of the metastable state and drive the transition is an interesting question. In the numerical works of Refs.~[\onlinecite{Pinettes2002,McClarty2014}], it was found rather difficult to reach the ground state by cooling through the metastable and by using local spin dynamics only.}. In their quantum analog~\cite{Kawamura1984,Chubokov1991,Zhitomirsky2000,Alicea2009,Coletta2013,Lee2014}, a quantum phase transition is driven by the competition between the classical energy and the energy correction arising from quantum fluctuations. However, these transitions are better attributed to both the parametric evolution of the ground state manifold and its associated fluctuations rather than due to fluctuations alone.

In this paper, we explore a different example of fluctuation-driven phase transition in the bilinear-biquadratic pyrochlore Heisenberg model:
\begin{align}
H &= \sum_{\langle ij \rangle} J\mathbf{S}_i\cdot\mathbf{S}_j+B(\mathbf{S}_i\cdot\mathbf{S}_j)^2\nonumber \\
& \equiv J_0\sum_{\langle ij \rangle} \cos\theta_\mathrm{B} \mathbf{S}_i\cdot\mathbf{S}_j+\frac{\sin\theta_\mathrm{B}}{S^2}(\mathbf{S}_i\cdot\mathbf{S}_j)^2.\label{eq:model}
\end{align}
$i$ and $j$ label the pyrochlore lattice sites, and the summation runs over all nearest-neighbor pairs. $J$ and $B$ are the bilinear and biquadratic exchange couplings, respectively. $S$ is the spin quantum number. $J \equiv J_0\cos(\theta_\mathrm{B})$, and $BS^2 \equiv J_0 \sin(\theta_\mathrm{B})$. We focus on the semi-classical limit, i.e. $S\gg 1 $ while $\theta_\mathrm{B}$ remains finite.

Considering antiferromagnetic systems with $J>0$, previous studies of the bilinear-biquadratic model in Eq.~\eqref{eq:model} have focused on the parameter space $-\pi/2<\theta_\mathrm{B}\le 0$, i.e.  $B\le0$~\cite{Reimers1992,Moessner1998a,Moessner1998b,Yamashita2000,Tchernyshyov2002,Penc2004,Shannon2010}. In this work, we focus on the opposite and largely unexplored parameter space with $0<\theta_\mathrm{B}<\pi/2$~\cite{*[{After submission of our manuscript, we were made aware (Nic Shannon, private communication) that a previous work (}] [{) that studies the problem of magnetic instability out of the fully polarized state of a pyrochlore magnetic system does consider the possibility of a positive biquadratic interaction $B$.}] Penc2007}. We find that, while the bilinear and biquadratic interactions admit a common, extensively degenerate classical ground state manifold, they produce different quantum fluctuations, which in turn select \emph{three} different states as one tunes the value $\theta_\mathrm{B}$ from $0$ to $\pi/2$. Thus, our results suggest a sequence of fluctuation-driven quantum phase transitions controlled by $\theta_\mathrm{B}$.

Our paper is organized as follows. In Sec.~\ref{sec:gs}, we chart the classical ground state manifold of the model in the parameter space $\theta_\mathrm{B}\in (0,\pi/2)$. In Sec.~\ref{sec:obd}, we study the order by disorder phenomenon, focusing on the effects of harmonic fluctuations and their role in the selection mechanism. We also discuss a plausible temperature versus $\theta_B$ phase diagram for the model. In Sec.~\ref{sec:outlook}, we provide an outlook toward potentially interesting directions to explore in the future as well as possible material realizations.


\section{Classical ground states}\label{sec:gs}

In this section, we explore the classical ground state manifold of the model defined by Eq.~\eqref{eq:model}. We begin with a discussion of the classical ground states of an isolated tetrahedron in Sec.~\ref{sec:tetrahedron}. In Sec.~\ref{sec:ice}, equipped with the result for an isolated tetrahedron, we show that an extensively large class of states, known as ``color ice states", belong to the classical ground state manifold of the full lattice model. In Sec.~\ref{sec:weathervane}, we demonstrate that color ice states support two-dimensional weathervane modes. These weathervane modes may give rise to nodal lines in the classical spin wave spectra, which we discuss in Sec.~\ref{sec:spinwave}.


\subsection{Classical ground state of a spin tetrahedron}\label{sec:tetrahedron}

\begin{figure}
\includegraphics[width = 0.8\columnwidth]{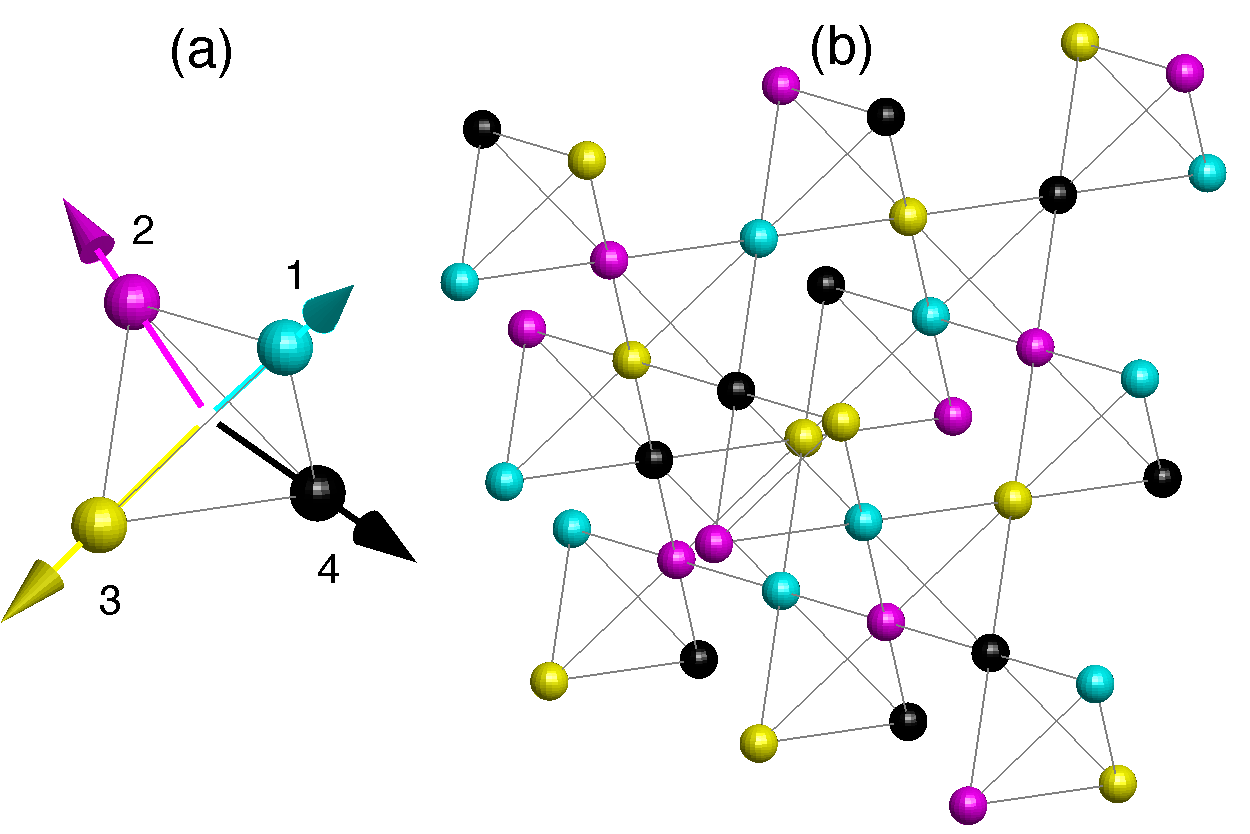}
\caption{(a) A classical ground state of an isolated tetrahedron. Arrows show the direction of the spins, $\mathbf{n}_1$, $\mathbf{n}_2$, $\mathbf{n}_3$, and $\mathbf{n}_4$, which are given in Eq.~\eqref{eq:spintetra}. Note that the specific directions of these spins, colored  cyan, magenta, yellow and black, shall be also respectively refered to as $\mathbf{n}_\mathrm{C}$, $\mathbf{n}_\mathrm{M}$, $\mathbf{n}_\mathrm{Y}$, $\mathbf{n}_\mathrm{K}$ in the discussion of Sec.~\ref{sec:ice} on color ice states (c.f. Eq.~\eqref{eq:colorcode}). (b) A color ice state. The color code indicates the spin orientation as per the convention in (a).}
\label{fig:tetra_ice}
\end{figure}

The pyrochlore lattice features a motif of corner-shared tetrahedra. It is customary to first study an isolated spin tetrahedron before embarking on a discussion of the full lattice model. To this end, we label the four classical spins at the corners of the tetrahedron as $1,2,3,4$. The interaction between a pair of spins takes the same form as in Eq.~\eqref{eq:model}. Replacing the operator $\mathbf{S}_i$ by the classical vector $S\mathbf{n}_i$, where $\mathbf{n}_i$ is a unit vector, we find the classical energy,
\begin{align}
E_\mathrm{tetra} = J_0S^2\sum_{i> j}\cos\theta_\mathrm{B} (\mathbf{n}_i\cdot\mathbf{n}_j)+\sin\theta_\mathrm{B} (\mathbf{n}_i\cdot\mathbf{n}_j)^2.
\label{eq:e_tetra}
\end{align}
Here $i,j$ run over the four sites of the tetrahedron.

A direct numerical minimization of Eq.~\eqref{eq:e_tetra} shows 
that the $\mathbf{n}_i$ in the classical ground state are given by,
\begin{align}
\mathbf{n}_1 &= \frac{1}{\sqrt{3}}\left(\begin{array}{c}
1 \\ 1\\ 1
\end{array}\right),\,
\mathbf{n}_2 =\frac{1}{\sqrt{3}}\left(\begin{array}{r}
-1 \\ -1\\ 1
\end{array}\right),\,\nonumber\\
\mathbf{n}_3 &=\frac{1}{\sqrt{3}}\left(\begin{array}{r}
1 \\ -1\\ -1
\end{array}\right),\,
\mathbf{n}_4 =\frac{1}{\sqrt{3}}\left(\begin{array}{r}
-1 \\ 1\\ -1
\end{array}\right).\label{eq:spintetra}
\end{align}

Here, each spin orientation is given in a column vector form. We see that the four spins form a tetrahedron in spin space, each pointing from the spin-tetrahedron centre to a corner (Fig.~\ref{fig:tetra_ice}a). In particular, $\mathbf{n}_i\cdot\mathbf{n}_j = -1/3$ between a pair of spins. A global rotation or inversion of the above state is also a classical ground state thanks to the $O(3)$ symmetry of Eq.~\eqref{eq:e_tetra}. We call the specific classical spin state described in Eq.~\eqref{eq:spintetra}, along with its global rotations and inversions, tetrahedral spin states (TSS).

For a general value of $\theta_B\in(0,\pi/2)$, we show rigorously that the TSS are in fact the global energy minima by using the following trick. We define a rank-2 traceless symmetric tensor,
\begin{align}
(\mathrm{T}_i)_{\alpha\beta} = n_{i\alpha}n_{i\beta}-\frac{1}{3}\delta_{\alpha\beta}.
\end{align}
Here $\alpha,\beta$ run over the three spin components $x,y,z$. $n_{i\alpha}$ is the $\alpha$ component of the unit vector $\mathbf{n}_i$. In particular, $\mathrm{tr}(\mathrm{T}^2_i) = 2/3$ and $\mathrm{tr}(\mathrm{T}_i\mathrm{T}_j) = (\mathbf{n}_i\cdot\mathbf{n}_j)^2-1/3$. Using these identities, we may rewrite Eq.~\eqref{eq:e_tetra} as
\begin{align}
\frac{E_\mathrm{tetra}}{J_0S^2} &= \frac{\cos\theta_\mathrm{B}}{2}(\sum_i\mathbf{n}_i)^2+\frac{\sin\theta_\mathrm{B}}{2}\mathrm{tr}(\sum_i\mathrm{T}_i)^2\nonumber\\
&-2\cos\theta_\mathrm{B}+\frac{2}{3}\sin\theta_\mathrm{B}.
\end{align}
Up to a constant independent of $\{\mathbf{n}_i\}$,  $E_\mathrm{tetra}$ is the sum of two squares. Hence, the energy can be minimized if one can find states for which $\sum_{i}\mathbf{n}_i=0$ and $\sum_{i}\mathrm{T}_i=0$ simultaneously. The first condition is the same as the familiar classical ground state condition for the bilinear Heisenberg model ($\theta_\mathrm{B}=0$) on a tetrahedron. Its solutions can be parametrized by two angles plus a global $O(3)$ operation~\cite{Moessner2001}. A few lines of algebra shows the second condition further constrains the solutions to TSS, namely Eq.~\eqref{eq:spintetra} and its rotations and inversions. Most importantly for the discussion that follows, any uniform (rigid) rotation of three of the $\mathbf{n}_j$ by an arbitrary angle $\phi$ about the fourth $\mathbf{n}_i$ ($i\ne j$) given by Eq.~\eqref{eq:spintetra} manifestly gives an energy minimum. Hence, we have proven that the TSS are the global minima of Eq.~\eqref{eq:e_tetra} for $0<\theta_\mathrm{B}< \pi/2$.


\subsection{Color ice states}\label{sec:ice}

Having determined the classical ground state manifold for an isolated tetrahedron, we move on to construct the classical ground states for the entire lattice. The classical energy of Eq.~\eqref{eq:model} is given by
\begin{align}
E = \sum_{\alpha}E_\alpha,
\label{eq:e_full}
\end{align} 
where $E_\alpha$ is the classical energy of a tetrahedron $\alpha$ (Eq.~\eqref{eq:e_tetra}). Since each $E_\alpha$ is minimized by a TSS, any four spins belonging to a tetrahedron must be in a TSS to minimize $E$. Said differently, the classical ground states of Eq.~\eqref{eq:model} must be the TSS assembled together.

To proceed, we first pick a reference tetrahedron in the lattice and put its four spins in a TSS. Without loss of generality, we take the orientation of these four spins defining our reference TSS to be identical to those given in Eq.~\eqref{eq:spintetra}. For reference sake, we assign four color labels, cyan (C), magenta (M), yellow (Y), and black (K) to these specific spin orientations (see Fig.~\ref{fig:tetra_ice}a)):
\begin{align}
\mathbf{n}_\mathrm{C} &\equiv \frac{1}{\sqrt{3}}\left(\begin{array}{r}
1 \\ 1\\ 1
\end{array}\right),\,
\mathbf{n}_\mathrm{M} \equiv \frac{1}{\sqrt{3}}\left(\begin{array}{r}
-1 \\ -1\\  1
\end{array}\right),\,\nonumber\\
\mathbf{n}_\mathrm{Y} &\equiv \frac{1}{\sqrt{3}}\left(\begin{array}{r}
1 \\ -1\\ -1
\end{array}\right),\,
\mathbf{n}_\mathrm{K} \equiv \frac{1}{\sqrt{3}}\left(\begin{array}{r}
-1 \\  1\\ -1
\end{array}\right).\label{eq:colorcode}
\end{align}

For the moment, we restrict the orientation of the remaining spins to this particular set, $\{\mathbf{n}_\mathrm{C},\mathbf{n}_\mathrm{M},\mathbf{n}_\mathrm{Y},\mathbf{n}_\mathrm{K}\}$. By construct, the four spins on any given tetrahedron are in a TSS if they take mutually different colors~\cite{Parameswaran2009}. Such a coloring rule is analogous to the familiar ice rule in Ising spin ice and is hence  known as ``color ice rule''. If every tetrahedron obeys the color ice rule, the energy of the full lattice is thereby minimal. Following the literature, we refer to the classical spin states in which every tetrahedron obeys the color ice rule as color ice states~\cite{Chern2011, Khemani2012, Chern2014}. An example of color ice states is illustrated in Fig.~\ref{fig:tetra_ice}b.

We thus have shown that the color ice states are part of the classical ground state manifold of Eq.~\eqref{eq:model}. By generalizing Pauling's estimate for Ising spin ice to color ice, one can show that the number of color ice states is approximately $(3/2)^{N/2}$, where $N$ is the number of lattice sites~\cite{Khemani2012}. Thus, the classical ground state manifold of Eq.~\eqref{eq:model} is exponentially large in the volume of the system.

We note that the above construction of the classical ground states of Eq.~\eqref{eq:model} is rather similar to the one presented in Ref.~[\onlinecite{Parameswaran2009}], where a different classical spin model arising from the Affleck-Kennedy-Lieb-Tasaki (AKLT) state on the pyrochlore lattice was studied. The authors of Ref.~[\onlinecite{Parameswaran2009}] also pointed out that color ice states may support zero-energy excitations. In the next section, we study these zero energy excitations in detail and elucidate their physical properties.


\subsection{Weathervane modes}\label{sec:weathervane}

\begin{figure}
\includegraphics[width=\columnwidth]{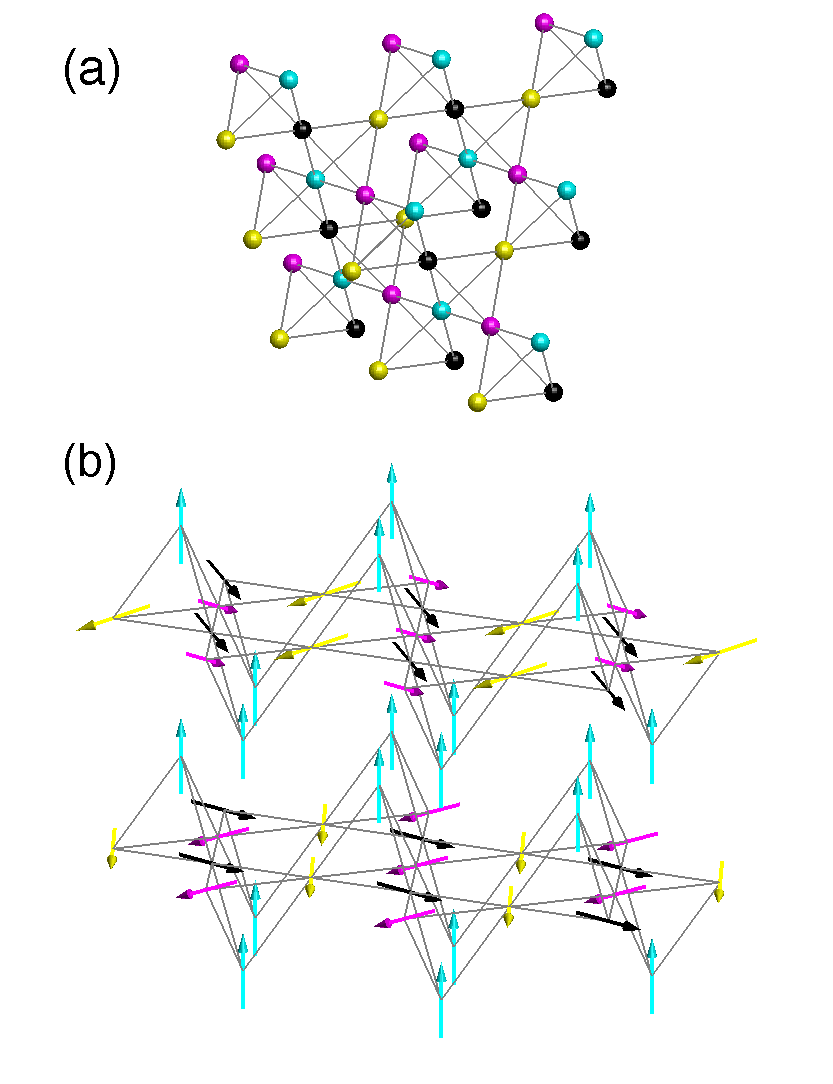}
\caption{(a) An all-in/all-out (AIAO) color ice state. (b) Top:  In the pyrochlore lattice, a kagome layer is sandwiched by two triangular layers. In the AIAO state shown in (a), the spins on the kagome layer are colored in magenta, yellow, and black, whereas spins on triangular layers are colored in cyan. Bottom: collectively rotating the spins on the kagome layer with respect to $\mathbf{n}_c$ (vertical-pointing cyan arrows) by an arbitrary weathervane angle $\phi_\mathrm{wv}$ preserves the TSS on every tetrahedron and thereby does not cost energy. Here, the kagome spins are rotated by $\phi_\mathrm{wv}=\pi/2$, which is \emph{not} a color ice state.}
\label{fig:aiao_wv}
\end{figure}

\begin{figure*}
\includegraphics[width=2\columnwidth]{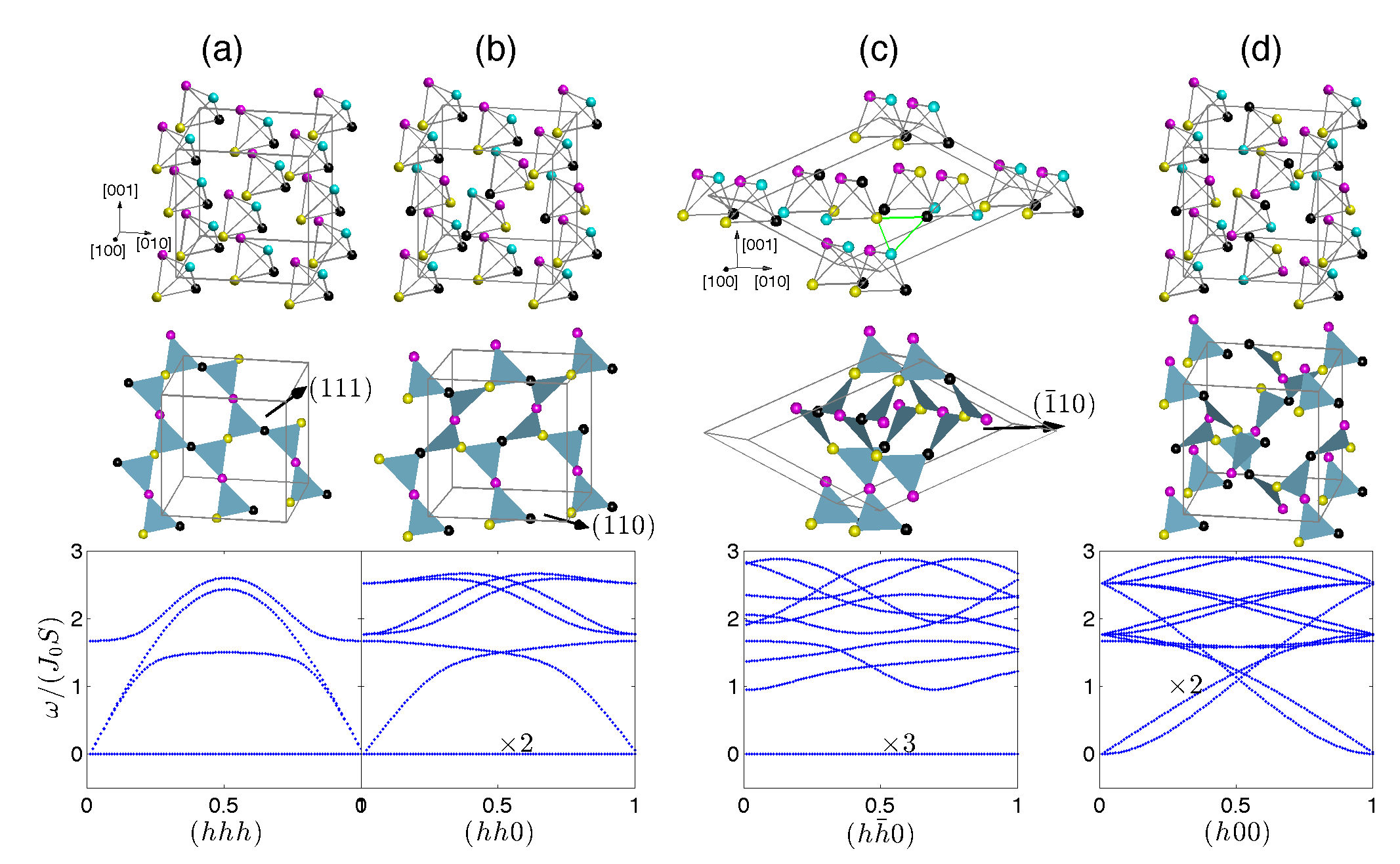}
\caption{(a) Top row: An all-in/all-out  state. Middle row: a weathervane membrane. There is no energy cost if spins belonging to the membrane rotate collectively with respect to $\mathbf{n}_\mathrm{C}$. The black arrow indicates the stacking direction of weathervane membranes. The grey box shows a conventional cubic unit cell. Bottom row: Classical spin wave dispersion along the stacking direction. Here we set $\theta_\mathrm{B} = 0.05\pi$ for all displayed spin wave calculation results. (b) A $\{100\}$ state. In this specific example, the modulation wave vector $\mathbf{q}=(001)$. In other words, the color (spin orientations) are invariant under translation along the $[100]$ and $[010]$ directions but alternating along the $[001]$ direction. The number $\times n$ in the panels showing the spin wave dispersion indicates the degeneracy of the branch. (c) A $\sqrt{3}\times\sqrt{3}$ state. This state is constructed by stacking the familiar $\sqrt{3}\times\sqrt{3}$ state in the kagome Heisenberg antiferromagnet~\cite{Harris1992,Chubukov1992,Reimers1993} along a $[111]$ direction. The color of the triangular (``apical'') sites is magenta. The grey box shows a unit cell.  One linking green triangle is being displayed as a visual aid to highlight a given kagome plane. Note the lowest three branches of the spin wave are dispersive along other directions in momentum space. In other words, they are not flat bands. (d) A color ice state which does not support a genuine weathervane mode as the weathervane membrane in fact covers all lattice sites.}
\label{fig:wv_spinwave}
\end{figure*}

We note that even though the above set of color ice states is exponentially large, such a set does not exhaust the classical ground state manifold of Eq.~\eqref{eq:model}~\cite{Parameswaran2009}. To illustrate this, we first consider an all-in/all-out (AIAO) long-range ordered state. In an AIAO state, every site of a given FCC sublattice has the same color, and, to fulfill the color ice rule, the four FCC sublattices must be assigned mutually different colors. Such a state preserves the translation symmetry of the pyrochlore lattice. An AIAO state is shown in Fig.~\ref{fig:aiao_wv}a.

We may also think of the pyrochlore lattice as alternately stacked triangular and kagome layers along the cubic $[111]$ direction. In an AIAO state, spins on the triangular layers are in the same color. In Fig.~\ref{fig:aiao_wv}b, the spins on triangular layers take the color cyan or, equivalently, the orientation $\mathbf{n}_\mathrm{C}$. We now focus on a given kagome layer and make the following observation: if we rotate the spins on the said layer by the same angle $\phi_\mathrm{wv}$ about the axis defined by $\mathbf{n}_C$ while keeping the rest of the spins unrotated, \emph{all} tetrahedra remain in TSS (Fig.~\ref{fig:aiao_wv}b). Therefore, the resulting state is a classical ground state of Eq.~\eqref{eq:model} according to our discussion in Sec.~\ref{sec:ice}. However, it is \emph{not} a color ice state if $\phi_\mathrm{wv}\neq 0$ or $\pm2\pi/3$ since the orientations of rotated spins are not in the set $\{\mathbf{n}_\mathrm{C},\mathbf{n}_\mathrm{M},\mathbf{n}_\mathrm{Y},\mathbf{n}_\mathrm{K}\}$\footnote{One may also perform mirror reflections on kagome spins at no energy cost. However, such operations do not produce new classical ground states.}.

We can perform the above collective rotations independently for each kagome layer, one layer at a time. Moreover, we may consider kagome layers stacked in other directions (for instance, $[\bar{1}\bar{1}1]$) and perform similar rotations. Therefore, starting from an AIAO state, we may obtain infinitely many classical ground states through such collective rotations, which are not color ice states except for special rotation angles. Such rotations are reminiscent of the weathervane modes in the classical kagome Heisenberg antiferromagnet~\cite{Chandra1991,Chalker1992,Harris1992,Chubukov1992,Sachdev1992,Huse1992,Ritchey1993}. By analogy, we therefore also refer to the aforementioned zero-energy-cost collective rotations as weathervane modes. In particular, we call the collective rotation angle $\phi_\mathrm{wv}$ of a weathervane mode the ``weathervane angle''. In the example presented in Fig.~\ref{fig:aiao_wv}b, the weathervane angle is $\phi_\mathrm{wv}=\pi/2$.

Many color ice states other than the AIAO state support weathervane modes. One may identify the weathervane modes supported by a color ice state through the following procedure. We pick a color, say cyan (C), and remove all the cyan sites in the pyrochlore lattice. Thanks to the color ice rule, each tetrahedron contains one and only one cyan site. The remaining sites thus form a network of corner-sharing triangles whose spins are colored magenta, yellow and black. This network may contain several connected components that are disconnected from each other. We then pick a connected component and rotate all the spins in this component with respect to $\mathbf{n}_\mathrm{C}$ by the same angle $\phi_\mathrm{wv}$. Such rotation preserves the TSS in every tetrahedron. Hence, we have found a weatherwave mode localized onto this component. We call the said connected component \emph{weathervane membrane} since its topology is that of a membrane. We show in Fig.~\ref{fig:wv_spinwave} a few more examples of weathervane membranes within various color ice states. In particular, we observe that the weathervane membranes possess diverse shapes.

An important question concerning the weathervane modes is how many such modes there are within a given color ice state. Our analysis of the AIAO states presented below shows that, for a crystal of linear dimension $L$ subject to periodic boundary conditions, the number of weathervane modes is of order $L$ (Fig.~\ref{fig:wv_spinwave}a). For the specific example shown in Fig.~\ref{fig:wv_spinwave}d, we find only one weathervane membrane covers all lattice sites. Hence, in this case, the weathervane mode coincides with a global spin rotation. In other words, this color ice state does not support a genuine weathervane mode. In Appendix~\ref{app:counting}, we prove the weathervane membrane must percolate through the lattice. Thus, the number of weathervane modes cannot be of order $L^3$, which would require localized modes and hence  contradict the percolation requirement. We also provide an argument that the counting is very unlikely to be $O(L^2)$. In addition, by randomly generating a large number of color ice states in a small system, we found that the AIAO states support the largest number of weathervane modes. We are therefore reasonably confident that the total counting of weathervane modes ranges from $0$ to $O(L)$.


\subsection{Spin waves about color ice states}\label{sec:spinwave}

In this section, we discuss the implications of weathervane modes on the classical spin waves in color ice states. The details of the spin wave calculations are presented in Sec.~\ref{sec:obd_formalism}. We note each weathervane mode give rise to a zero mode in the spin wave spectrum. In the AIAO state shown in Fig.~\ref{fig:wv_spinwave}a, the weathervane membranes are stacked along a $[111]$ direction. Thus, the spin wave spectrum must contains nodal lines along the $\{111\}$ directions as shown in Fig.~\ref{fig:wv_spinwave}a. Similarly, the spin wave spectra in the color ice states shown in Fig.~\ref{fig:wv_spinwave}b and \ref{fig:wv_spinwave}c contain nodal lines as well. However, since the color ice state in Fig.~\ref{fig:wv_spinwave}d  does not support non-trivial weathervane modes, the spin wave spectrum contains no nodal lines as shown in the bottom panel of Fig.~\ref{fig:wv_spinwave}d. Finally, we remark that the nodal lines in the spin wave spectra are reminiscent of the spin wave spectra of the FCC Heisenberg antiferromagnet~\cite{Haar1962,Lines1963}.


\section{Order by disorder}\label{sec:obd}

In the previous section, we showed that the classical ground state manifold of Eq.~\eqref{eq:model} possesses a rich structure. The color ice states may be visualized as a set of discrete points. Starting from a color ice state, a weathervane mode generates  a one-dimensional sub-manifold, whose topology is that of a circle. Each point on this circle corresponds to a different value of weathervane angle $\phi_\mathrm{wv}$, with $\phi_\mathrm{wv}=0,\pm2\pi/3$ being the color ice states. Therefore, the classical ground state manifold is a hybrid of continuous and discrete structures.

In this section, we study the order by disorder phenomenon in the classical ground state manifold of Eq.~\eqref{eq:model} within a harmonic approximation. In Sec.~\ref{sec:obd_formalism}, we briefly review the formalism. In Sec.~\ref{sec:obd_wv}, we study the thermal and quantum selection within the one-dimensional sub-manifold generated by a weathervane mode.  By investigating several one-dimensional sub-manifolds, we find that the entropy maxima are invariably two-fold dgenerate. One maximum is located at a color ice state, and the other is located at a non-color-ice state, which is related to the said color ice state by a weathervane mode with $\phi_\mathrm{wv}=\pi$. Similarly, the energy minima are also two-fold degenerate, and the minima are located at the same states. Furthermore, we point out that such two-fold degeneracy results from a special property of the dynamical matrix. In Sec.~\ref{sec:obd_ice}, we perform a restricted search for the maximal entropy and minimal energy states among the color ice states. We find \emph{two} different maximal entropy states for different values of the model parameter $\theta_\mathrm{B}$. While the minimal zero point energy states coincide with the maximal entropy states for large and small values of $\theta_\mathrm{B}$, a \emph{third} color ice state arises as the minimal energy state in a small window of $\theta_\mathrm{B}$. In Sec.~\ref{sec:obd_phases}, we discuss the implications of our finding on the phases of the model Eq.~\eqref{eq:model}. In particular, we argue that our results suggest two unusual fluctuation-driven phase transitions in an extensively degenerate classical ground state manifold.


\subsection{Formalism}\label{sec:obd_formalism}

To begin, we derive the dynamical matrix for the harmonic fluctuations near a classical ground state. We replace the spin operator $\mathbf{S}_i$ by classical vector $S\mathbf{n}_i$ and rewrite the unit vector $\mathbf{n}_i$ as,
\begin{align}
\mathbf{n}_i = \hat{z}_i(1-\frac{u^2_i}{2})+\mathbf{u}_i\sqrt{1-\frac{{u_i}^2}{4}}.\label{eq:marsaglia}
\end{align}
$\hat{z}_i$ is the unit vector that parametrizes the orientation of the spin $i$ in the classical ground state. Two-dimensional vector $\mathbf{u}_i\perp\hat{z}_i$ parametrizes the deviation from the classical ground state with $u_i\le2$. In particular, we recover $\mathbf{n}_i=\hat{z}_i$ when $\mathbf{u}_i=0$. Importantly, the Jacobian of the transformation in Eq.~\eqref{eq:marsaglia} is 1~\cite{*[{We note the transformation Eq.~\eqref{eq:marsaglia} is reminiscent of the Holstein-Primakoff transformation for quantum spins. To our knowledge, its explicit form was first proposed in }] [{ in the context of geometric probability.}] Marsaglia1972}.
 
We plug Eq.~\eqref{eq:marsaglia} into the expression for classical energy Eq.~\eqref{eq:e_full} and expand $E$ up to second order in $u_i$, $E \approx E_0+E^{(2)}$. The classical ground state energy is $E_0 \equiv NJ_0S^2(-\cos\theta_\mathrm{B}+\sin\theta_\mathrm{B}/3)$, where $N$ is the number of sites. The quadratic piece is
\begin{align}
E^{(2)} &\equiv \frac{K_0}{2}\sum_{i}u^2_i+K_1\sum_{\langle ij\rangle}\mathbf{u}_i\cdot\mathbf{u}_j\nonumber\\
&+K_2\sum_{\langle ij\rangle}(\mathbf{u}_i\cdot\hat{z}_j)(\mathbf{u}_j\cdot\hat{z}_i),
\label{eq:quadratic}
\end{align}
 where $K_0 = 2J_0S^2(\cos\theta_\mathrm{B}+2\sin\theta_\mathrm{B})$, $K_1 = J_0S^2(\cos\theta_\mathrm{B}-2/3\sin\theta_\mathrm{B})$ and $K_2 = 2J_0S^2\sin\theta_\mathrm{B}$.

To proceed, we introduce explicit frames $\mathbf{u}_i = u_{ix}\hat{x_i}+u_{iy}\hat{y_i}$. The choice for $\{\hat{x}_i,\hat{y}_i\}$ is arbitrary as long as $\{\hat{x}_{i},\hat{y}_i,\hat{z}_i\}$ form a right-handed orthonormal basis. All physical observables are independent of such a choice. $E^{(2)}$ is then recast as a quadratic form of $u_{ix},u_{iy}$,
\begin{subequations}
\begin{align}
E^{(2)} = \frac{1}{2}\sum_{i\alpha,j\beta}\mathrm{M}_{i\alpha,j\beta}u_{i\alpha}u_{j\beta}.
\end{align}
Here the summation of $\alpha,\beta$ runs over $x$ and $y$. The $2N \times 2N$ dynamical matrix $\mathrm{M}$ encodes the information about the harmonic fluctuations near a classical ground state. The diagonal elements of $\mathrm{M}$ are given by
\begin{align}
\mathrm{M}_{ix,ix} = \mathrm{M}_{iy,iy} = K_0,\quad
\mathrm{M}_{iy,ix} = \mathrm{M}_{ix,iy} =0.
\end{align}
The off-diagonal elements are non-zero if and only if $i,j$ are nearest-neighbors, with
\begin{align}
\mathrm{M}_{i\alpha,j\beta} = K_1 g_{i\alpha,j\beta}+K_2 g_{i\alpha,jz}g_{iz,j\beta}.
\label{eq:mentry}
\end{align}
\end{subequations}
Here, $g_{i\alpha,j\beta}\equiv\hat{\alpha}_i\cdot\hat{\beta}_j$ with $\alpha,\beta$ taking the label $x$ or $y$. Likewise, $g_{i\alpha,jz}\equiv\hat{\alpha}_i\cdot\hat{z}_j$, and $g_{iz,j\beta}\equiv\hat{z}_i\cdot\hat{\beta}_j$.

\emph{Thermal selection.}  Once the dynamical matrix $\mathrm{M}$ is known, one can compute the entropy from classical thermal harmonic fluctuations using
\begin{align}
S_\mathrm{cl} = -\frac{k_\mathrm{B}}{2}\mathrm{tr}'\ln\mathrm{M}.\label{eq:cl_entropy}
\end{align}
The thermal fluctuations favor the classical ground state with maximal $S_\mathrm{cl}$. Note that we exclude the zero modes of $\mathrm{M}$ in the trace and hence the prime on $\mathrm{tr}$ in Eq.~\eqref{eq:cl_entropy}~\footnote{In Eq.~\eqref{eq:cl_entropy}, we have neglected the contribution from zero modes. The weathervane modes give rise to zero modes (Sec.~\ref{sec:spinwave}). However, the number of zero modes do not exceed $O(L)$ in a periodic system of linear dimension $L$ (Sec.~\ref{sec:weathervane}). Therefore, $S_\mathrm{cl}$ is dominated by the contribution from modes of finite frequency, whose number is of order $L^3$. Yet, it is important to note that the covariance matrix, $\mathrm{M}^{-1}$, is not well defined due to $O(L)$ zero modes.}.

\emph{Quantum selection.} Given $\mathrm{M}$, one can also compute the classical spin wave spectrum by solving the following equation of motion,
\begin{align}
\dot{u}_{i\alpha} = J_0S\sum_{j\beta}(\upeta\mathrm{M})_{i\alpha,j\beta}u_{j\beta},
\label{eq:spindyn}
\end{align}
where $\upeta$ is the $2N\times 2N$ skew-symmetric matrix: $\upeta_{i\alpha,j\beta}=\eta_{\alpha\beta}\delta_{ij}$, where $\eta_{xx}=\eta_{yy}=0$, and $\eta_{xy} = \eta_{yx} = -1$. The classical spin wave spectrum is then found by diagonalizing $i\upeta\mathrm{M}$. It can be shown that $i\upeta\mathrm{M}$ is diagonalizable if $\mathrm{M}$ is semi-positive-definite~\cite{Blaizot1986}. Moreover, the eigenvalues of $i\upeta\mathrm{M}$ come in pairs: $\pm\omega_1,\pm\omega_2\cdots \pm\omega_N$, where $\omega_\lambda$ are real and non-negative. The eigenfrequencies of spin wave modes are simply given by $\{\omega_\lambda\}$.

Upon quantization, each spin wave mode $\lambda$ contributes a zero point energy $\omega_\lambda/2$. The total energy at zero temperature is given by,
\begin{align}
E_\mathrm{qt} = \frac{1}{2}\sum_{\lambda}\omega_\lambda,
\label{eq:zpe}
\end{align}
where the summation runs over all spin wave modes $\lambda$. The quantum fluctuations select the state with minimal $E_\mathrm{qt}$ as the ground state of the Hamiltonian.

\emph{Free energy at finite temperature.} The selection effects of harmonic thermal fluctuations and harmonic quantum fluctuations may be treated on equal footing by considering the free energy at finite temperature $T$: 
\begin{align}
F = \sideset{}{'}\sum_{\lambda}\frac{\omega_\lambda}{2}+k_\mathrm{B}T\ln(1-e^{-\beta\omega_\lambda}),
\label{eq:free_energy}
\end{align}
where we exclude the zero modes ($\omega_\lambda=0$) from the summation. $\beta \equiv 1/(k_\mathrm{B}T)$. In the limit of $T\to\infty$,
\begin{align}
F &\to k_\mathrm{B}T\sum_{\lambda}\ln\omega_\lambda = \frac{k_\mathrm{B}T}{2}\ln[(-)^N\det(i\upeta\mathrm{M})]\nonumber\\
& = \frac{k_\mathrm{B}T}{2}\mathrm{tr}\ln\mathrm{M}=-TS_\mathrm{cl},
\end{align} 
which recovers the classical result of Eq.~\eqref{eq:cl_entropy}. The second equality follows from the spectral property of $i\upeta\mathrm{M}$ that its eigenvalues come in pairs, and the third equality follows from $\det(i\upeta) = (-1)^N$. We have assumed above that $i\upeta\mathrm{M}$ contains no zero eigenvalues for the sake of simplicity (they can be removed by an infinitesimally small 
straggered magnetic field).


\subsection{Selection of weathervane modes}\label{sec:obd_wv}

\begin{figure}
\includegraphics[width=\columnwidth]{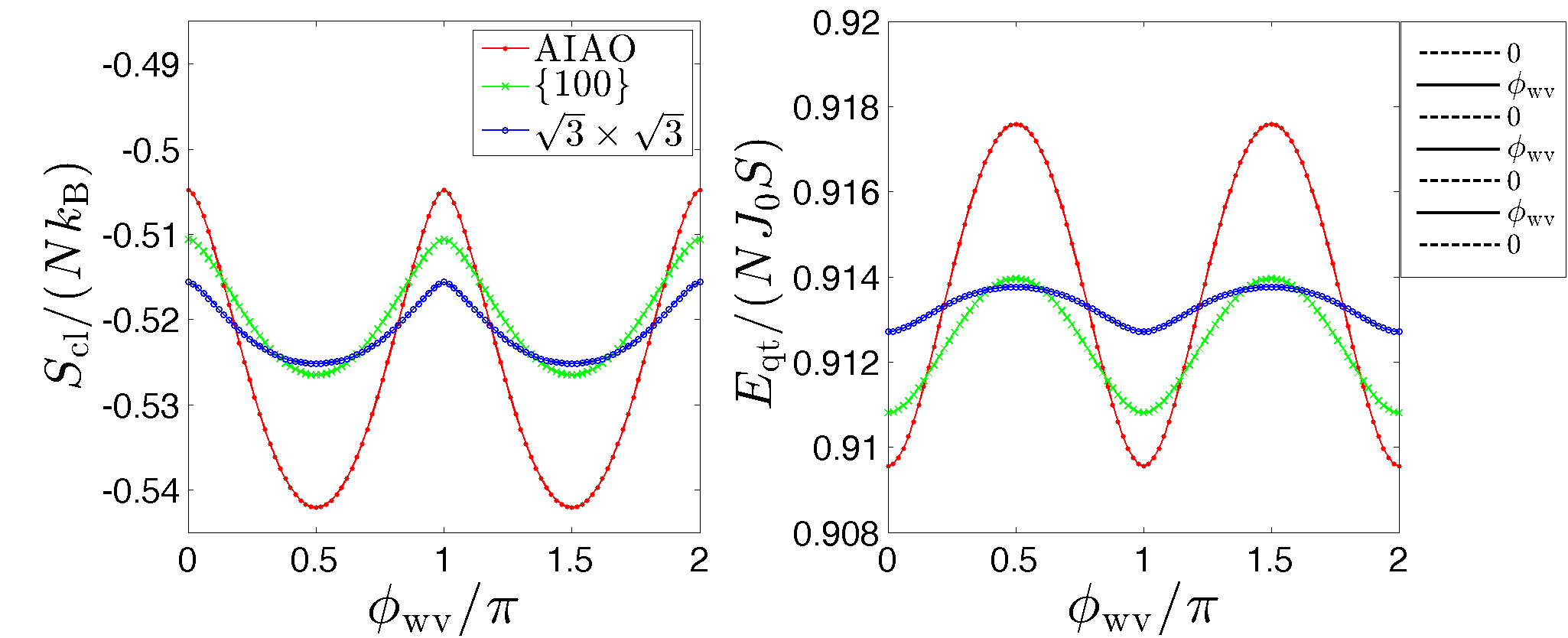}
\caption{Left: Classical entropy per site as a function of the weathervane mode rotation angle $\phi_\mathrm{wv}$ about an AIAO (red dots), a $\{100\}$(green crosses), and a  $\sqrt{3}\times\sqrt{3}$ (blue open circles) state. $\theta_\mathrm{B} = 0.05\pi$ in our calculation. Systems of $30\times 30\times 30$ magnetic unit cells are used. Right: Similar to the left panel but for the zero point energy per site. Outside the right panel: schematic illustration of the calculation setup. Spins on even weathervane membranes (dashed lines) are fixed whereas the spins on even weathervane membranes (solid lines) are rotated by the same angle $\phi_\mathrm{wv}$.
}
\label{fig:wv_select}
\end{figure}

Equipped with the formalism presented in Sec.~\ref{sec:obd_formalism}, we can now study the thermal and quantum selection in the one-dimensional sub-manifold generated by a weathervane mode. Given that each color ice state may support infinitely many weathervane mode, and that the number of color ice states is an exponential function of the system size, we cannot exhaust all such sub-manifolds. Instead, we investigate several simple and typical examples and draw tentative conclusions from them.

We consider weathervane modes in an AIAO state, a $\{100\}$ state, and a $\sqrt{3}\times\sqrt{3}$ state (Fig.~\ref{fig:wv_spinwave}), in which the spins rotate with respect to $\mathbf{n}_\mathrm{C}$. In all three states, the weathervane membranes are stacked (hence the ``even''/``odd'' label later) on one another. For the sake of simplicity, we fix the spins on even weathervane membranes and rotate the spins on odd membranes by the same angle $\phi_\mathrm{wv}$ (Fig.~\ref{fig:wv_select}, outside right panel). To avoid computational issues arising from the zero modes (Sec.~\ref{sec:obd_formalism}), we impose anti-periodic boundary conditions in two directions and periodic boundary condition in the third. We consider systems of $L\times L\times L$ magnetic unit cells. We set $L=30$ and find no visible change if we increase $L$.

The results are summarized in Fig.~\ref{fig:wv_select}. We set $\theta_\mathrm{B}=0.05\pi$ for illustration purpose. The behavior of $S_\mathrm{cl}$ or $E_\mathrm{qt}$ as function of $\phi_\mathrm{wv}$ is qualitatively the same for other choice of the model parameter $\theta_\mathrm{B}$. Despite the diverse shapes of the weathervane membranes, in all three cases, the entropy maxima are at $\phi_\mathrm{wv}=0$ and $\pi$. Likewise, the zero point energy minima are also at $\phi_\mathrm{wv}=0$ and $\pi$. Thus, both thermal and quantum fluctuations select the same states in a given sub-manifold. Note $\phi_\mathrm{wv}=0$ are color ice states whereas $\phi_\mathrm{wv}=\pi$ are not. Furthermore, $\phi_\mathrm{wv}=2\pi/3$ and $4\pi/3$, which also correspond to color ice states, are neither local entropy maxima nor local 
(zero point) energy minima. Hence, not all color ice states are stable against a weathervane mode. Finally, the energetic stability of the AIAO, $\{100\}$, and $\sqrt{3}\times\sqrt{3}$ states against small $\phi_\mathrm{wv}$ implies that the nodal lines in the classical spin wave spectra would acquire dispersion at zero temperature once quantum fluctuations are fully taken into account (Sec.~\ref{sec:spinwave}).

The $\phi_\mathrm{wv}=0$ and $\phi_\mathrm{wv}=\pi$ states are degenerate both in classical entropy and in zero point energy. Such degeneracy is not coincidental. In Appendix \ref{app:dyna_sym}, we prove that the dynamical matrices $\mathrm{M}$ in two classical ground states related by a $\phi_\mathrm{wv}=\pi$ rotation in a weathervane membrane are identical up to a local gauge transformation. As a corollary, the two ground states have the same classical entropy (Eq.~\eqref{eq:cl_entropy}) and the same zero point energy (Eq.~\eqref{eq:zpe}). For later convenience, We refer as  color-companion (CC) a classical ground state that is obtained from an initial color ice state through a weathervane mode with a weatherwave angle $\phi_\mathrm{wv} = \pi$. Note a color ice state may support many weathervane modes, each weathervane mode generating a CC of that color ice state. For instance, an AIAO state supports infinitely many weathervane modes, each localized onto a kagome layer of the pyrochlore lattice (Sec.~\ref{sec:weathervane}). One may pick a specific kagome layer and collectively rotate the spins in that layer by an angle $\phi_\mathrm{wv}=\pi$, whereby obtaining a CC of the original AIAO state. Since there are many kagome layers and one can rotate them one by one, the AIAO state has infinitely many CCs.


\subsection{Selection of color ice states}\label{sec:obd_ice}

\begin{figure}
\includegraphics[width=\columnwidth]{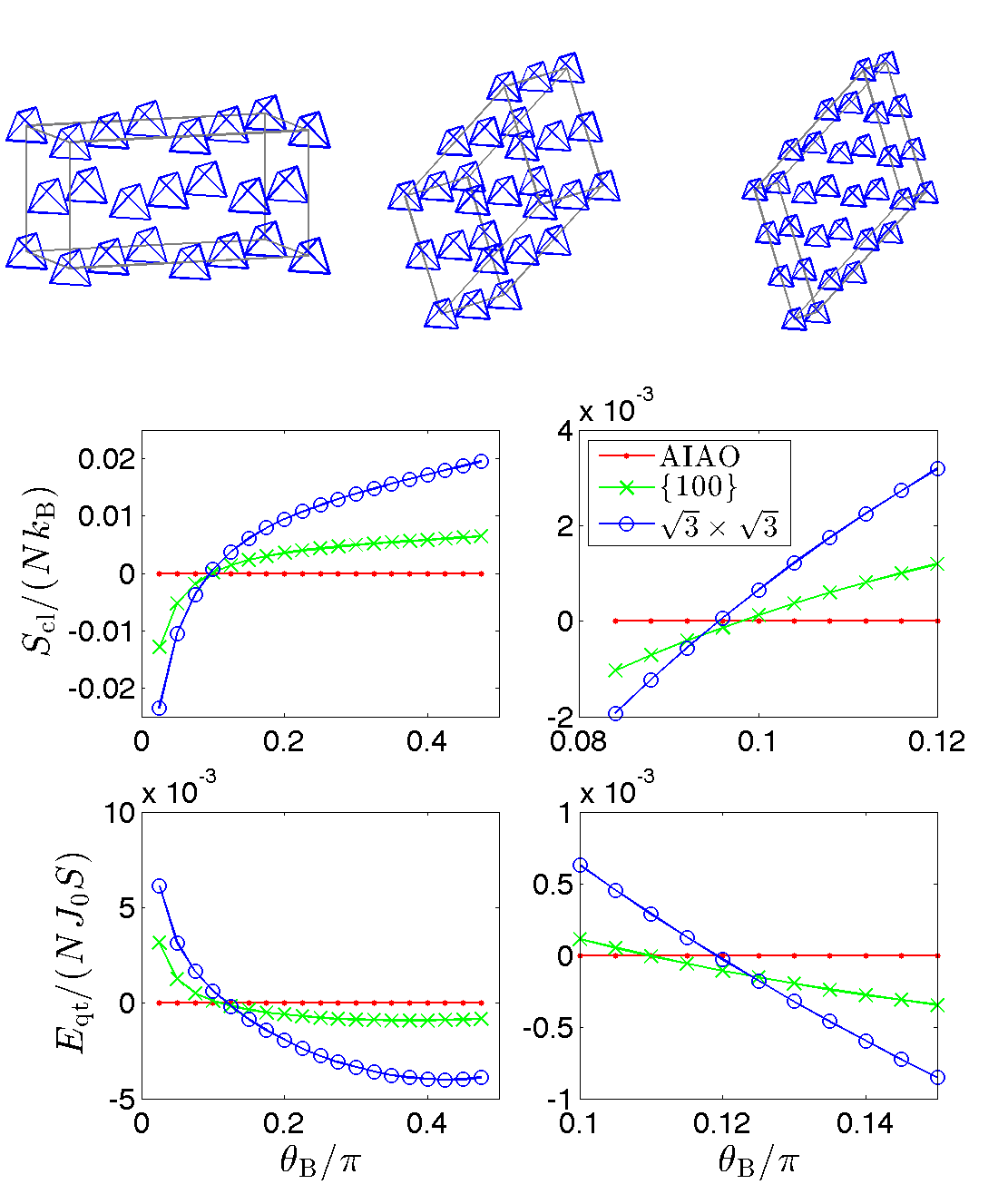}
\caption{Top: Various supercells used in the search for maximal entropy states. From left to right: cubic $2\times1\times1$ supercell, rhombohedral $2\times2\times2$ supercell, and rhombohedral $3\times3\times1$ supercell. Grey boxes highlight the boundary of the supercells. Middle left: entropy per site of the AIAO (red dots), $\{100\}$ (green crosses), and $\sqrt{3}\times\sqrt{3}$ states (blue open circles) as a function of $\theta_\mathrm{B}\in (0,\pi/2)$. The entropy of AIAO states is subtracted for better discernibility. Middle right: the entropy as a function of $\theta_\mathrm{B}$ in a smaller interval $\theta_\mathrm{B}\in (0.08\pi,0.12\pi]$. Bottom: similar to the middle panels but for the zero point energy per site. We use a $30\times 30\times 30$ grid of $2\times2\times2$ rhombohedral supercells in calculating the entropy or energy of AIAO states and $\{100\}$ states and a $20\times 20\times 60$ grid of $3\times3\times1$  rhombohedral supercells for $\sqrt{3}\times\sqrt{3}$ states. Our choice of system size ensures the number of sites is the same.}
\label{fig:full_search}
\end{figure}

\begin{figure}
\includegraphics[width=\columnwidth]{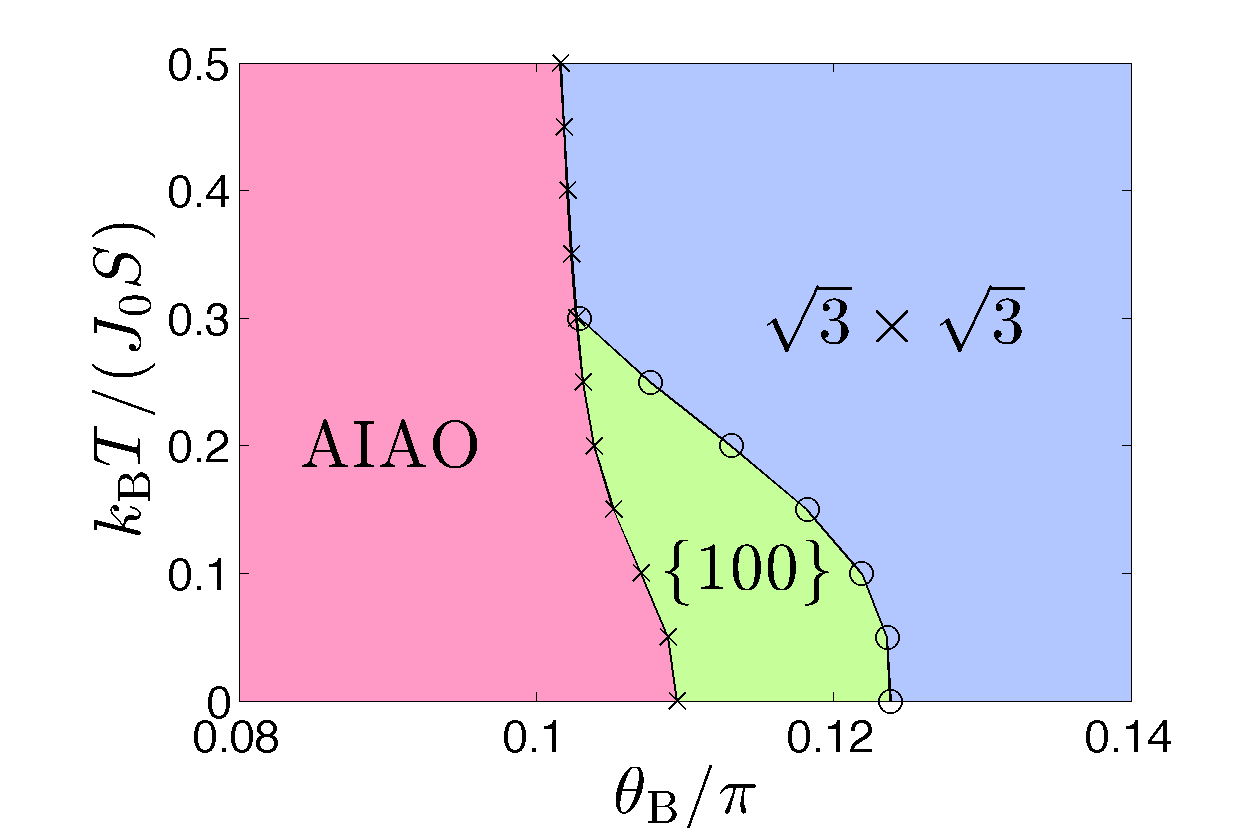}
\caption{Color ice states with minimal free energy at finite temperature. Regions with different minimal free energy states are colored differently. Crosses and open circles mark regional boundaries. The calculation set up is the same as in Fig.~\ref{fig:full_search}.}
\label{fig:free_energy}
\end{figure}

In the previous subsection, we found that within the one-dimensional sub-manifold generated by a weathervane mode, the maximal entropy states are a color ice state and a CC. Even though we explored only three one-dimensional sub-manifolds, we hypothesize that this is true in general, and speculate that the maximal entropy states in the entire classical ground state manifold are a color ice state and its CCs. In particular, since the classical entropy of a color ice state and its CCs is degenerate, we can simply search for the maximal entropy states among the color ice states. The same reasoning applies to the search for the minimal quantum zero point energy states, and hence we consider only the color ice states as well.

The problem remains a formidable one even after such a simplification because the number of color ice states is an exponential function of system size (Sec.~\ref{sec:ice}). We are therefore forced to consider a limited set of candidate color ice states. Yet, we postulate that both the maximal entropy states and the minimal energy states are commensurate with a small magnetic supercell. We then perform a brute-force search among all commensurate color ice states.

We use three supercells: the $2\times 1\times 1$ cubic supercell made by stacking crystallographic unit cells of the pyrochlore lattice; the $2\times 2\times 2$ rhombohedral supercell made by stacking the primitive unit cells of the pyrochlore lattice; the $3\times 3\times 1$ rhombohedral supercell (Fig.~\ref{fig:full_search}, top panel). To compute entropy (energy), we stack the supercells in all three directions and diagonalize the matrix $\mathrm{M}$ ($i\upeta\mathrm{M}$) for this large system (Sec.~\ref{sec:obd_formalism}).

The results are summarized in Fig.~\ref{fig:full_search}. We find the maximal entropy states are the AIAO states when $\theta_\mathrm{B}<0.0916\pi$ and the $\sqrt{3}\times\sqrt{3}$ states when $\theta_\mathrm{B}>0.0916\pi$. While there is close competition between the $\{100\}$ states and the $\sqrt{3}\times\sqrt{3}$ states near $\theta_\mathrm{B}=0.09\pi$, the latter becomes the one with largest entropy once the AIAO states have lost stability. 

The zero point energy shows similar behavior. The minimal energy state are the AIAO states for $\theta_\mathrm{B}<0.110\pi$ and the $\sqrt{3}\times\sqrt{3}$ states for $\theta_\mathrm{B}>0.124\pi$. However, different from the entropy case, the $\{100\}$ states become the minimal energy states in a small window $0.110\pi<\theta_\mathrm{B}<0.124\pi$. 

While both the classical thermal fluctuations and zero temperature quantum fluctuations favor the same color ice states at large and small $\theta_\mathrm{B}$, they differ in a small $\theta_\mathrm{B}$ interval. To gain a better understanding of the selection effect when both quantum and thermal fluctuations are present, we compare the free energy 
(c.f. Eq.~\eqref{eq:free_energy}) of the three competing color ice states. The results are summarized in Fig.~\ref{fig:free_energy}. We find that, the $\theta_\mathrm{B}$ window of stability for the $\{100\}$ states diminishes as the temperature $T$ increases from 0, and eventually disappears when $T/(J_0S) \gtrsim 0.3$. While we caution that this calculation is carried out at the harmoic level, it is interesting that the quantum calculation shows a collapse of the $\{100\}$ phase at sufficiently high temperature, giving only two competing phases (AIAO and the $\sqrt{3}\times\sqrt{3}$) as in the classical entropy calculation.

\subsection{Semiclassical phase diagram}\label{sec:obd_phases}

\begin{figure}
\includegraphics[width=0.8\columnwidth]{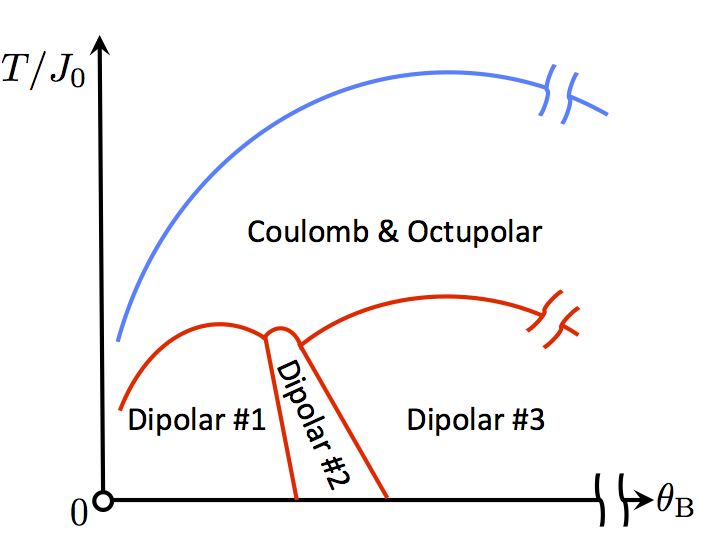}
\caption{Sketch of the speculated phase diagram of Eq.~\ref{eq:model} in the semi-classical limit ($S\gg 1$). The phase boundary curves are mere guide for eye. $\theta_\mathrm{B}=0$ corresponds to the bilinear Heisenberg model (marked as open circle), where the physics is very different from the bilinear-biquadratic model discussed here. See Sec.~\ref{sec:outlook} for discussion on this limit.}
\label{fig:phasediagram}
\end{figure}

In this subsection, we discuss the implications of our finding in Sec.~\ref{sec:obd_ice} on the phase diagram of the model Eq.~\eqref{eq:model} in the semiclassical limit. A sketch of the speculated phase diagram is presented in Fig.~\ref{fig:phasediagram}.

We first consider $T=0$, where the quantum fluctuations dominate. Recall that the zero point energy of a color ice state and its CCs is degenerate. Thus, when $\theta_\mathrm{B}<0.110\pi$, harmonic quantum fluctuations in fact not only select the AIAO states but also their CCs. However, the degeneracy among this set of states $\mathcal{S}$, namely the AIAO states and their CCs, is accidental as it results from a peculiarity of the quadratic approximation (Sec.~\ref{sec:obd_wv}). At a higher-order approximation, non-harmonic quantum fluctuations would lift the accidental degeneracy and select a unique member state (up to the global symmetries of Eq.~\eqref{eq:model}) from $\mathcal{S}$. Since this state exhibits long-range dipolar magnetic order, we denote it as Dipolar \#1.  By the same reasoning, quantum fluctuations select a second (denoted as Dipolar \#2) and a third ground state (denoted as Dipolar \#3) in the parameter window $\theta_\mathrm{B}\in (0.110\pi,0.124\pi)$ and $\theta_\mathrm{B}\in (0.124\pi,0.5\pi)$, respectively. Specifically, Dipolar \#2 state is either $\{100\}$ state or its CCs, and Dipolar \#3 state is either $\sqrt{3}\times\sqrt{3}$ or its CCs. In short, our results suggest two fluctuation-driven phase transitions at $T=0$.

We must caution that our finding is based on a restricted search. It is possible that there may exist a state outside our searched space whose energy is minimal for all $0<\theta_\mathrm{B}<\pi/2$, whereby evading the phase transition discussed here. We believe this is unlikely given the high symmetry of the AIAO states and the fact that the AIAO states have minimal energy at small $\theta_\mathrm{B}$ for the three supercells investigated. Yet, the exact nature of the minimal energy states for larger $\theta_\mathrm{B}$ remains less certain at this time.

When $T>0$, the thermal fluctuations set in. Since all three ground states are long-range ordered and the system is three-dimensional, they are stable at sufficiently low but finite temperature. Our finite temperature free energy calculation (Fig.~\ref{fig:free_energy}) shows the two critical $\theta_\mathrm{B}$s decreases as $T$ increases. Each of the three long-range ordered state ultimately disappear above its respective critical temperature. Even though the free energy calculation shows that two phase transitions may merge into one at $T/(J_0S) \gtrsim 0.3$, it seems unlikely to occur as it may have been well above the critical temperature of the Dipolar \#2 state.

The dipolar long-range orders disappear above their critical temperature. If the temperature remain sufficiently smaller than the biquadratic interaction energy scale $(J_0S^2\sin\theta_\mathrm{B})$, each tetrahedron must be in a TSS. We speculate that the system is then thermally fluctuating among the color ice states, implying that it is in a Coulomb phase~\cite{Khemani2012,Chern2014}. Moreover, the spins in color states are restricted to four symmetric directions (Fig.~\ref{fig:tetra_ice}a). As a result, the system also exhibits an octupolar magnetic order~\cite{Zhitomirsky2008}. We note that the thermal fluctuations dominate in this temperature regime. Even though our calculation (Fig.~\ref{fig:full_search}) shows the AIAO and the $\sqrt{3}\times\sqrt{3}$ states are respectively the color ice states with maximal classical entropy for small and large $\theta_\mathrm{B}$, thermal fluctuations alone are unlikely to stabilize any dipolar long-range-ordered state as the entropy gain would be too small to compete with that of a Coulomb phase~\cite{Moessner1998b,Shannon2010}. In particular, we do not anticipate any dipolar long range order in the strictly classical model. Yet, the thermal fluctuations should be able to induce different types of short range orders in their respective model parameter space.

Finally, as the temperature increases further, the octupolar magnetic order melts through a thermodynamic phase transition into either a Coulomb phase or a trivial paramagnetic phase, a subtlety that we are unable to expand much on here. Note that a true phase transition, not merely a crossover, must occur since the octupolar magnetic order breaks spin rotational and time reversal symmetries whereas the high temperature phase respects all symmetries.


\section{Outlook}\label{sec:outlook}

We stress that the phase diagram illustrated in Fig.~\ref{fig:phasediagram} is speculative. As pointed out in Sec.~\ref{sec:obd_ice}, the exact nature of the minimal energy state for large $\theta_\mathrm{B}$ is less certain given that our result is based on a restricted search. Simulated annealing would be necessary to resolve this issue. Furthermore, a classical Monte Carlo study is required to put the postulated octupolar Coulomb phase on a firmer ground. Similar to the kagome Heisenberg antiferromagnet and the hyperkagome Heisenberg antiferromagnet, Eq.~\eqref{eq:model} possess extensively degenerate non-collinear ground states. Performing Monte Carlo simulation of such systems is non-trivial and would require a specially tailored algorithm~\cite{Huse1992,Reimers1993,Petrenko2000,Hopkinson2007,Zhitomirsky2008,Schnabel2012,Chern2013}. Notwithstanding this difficulty, a classical Monte Carlo investigation of Eq.~(\ref{eq:model}) would likely prove interesting.

Throughout this paper, we consider the semi-classical limit ($S\gg 1$, $BS^2\sim J$) of the model Eq.~\ref{eq:model}. Specifically, our calculation of the zero point energy is within the quadratic approximation, or equivalently to the order of $1/S$. At this order, there is a \emph{partial} lifting of energy degeneracy among the classical ground states. We have argued that the non-harmonic fluctuations would fully lift the 
remaining accidental degeneracy between color ice states and the color companions. To determine the actual ground state, one would need to go to higher order in $1/S$~\cite{Chubukov1992, Hizi2009,Chernyshev2014, Chernyshev2015}. In the opposite quantum regime ($S\sim 1$), more possible ground states, including quantum spin liquid, valence bond solids, and quantum multipolar magnetic order may appear, a matter that would be interesting to explore as well.

Another subtlety arises when we consider the limit $\theta_\mathrm{B}\to 0$.
This limit corresponds to the familiar bilinear antiferromagnetic Heisenberg model on pyrochlore lattice, which has a much larger classical ground state manifold~\cite{Moessner1998a,Moessner1998b}. To the order of $1/S$, the quantum fluctuations partially lift the degeneracy and select a family of collinear states~\cite{Henley2006,Hizi2006,Hizi2009}. Once $\theta_\mathrm{B}>0$, the collinear states become metastable saddle points in the classical energy landscape. For sufficiently small but finite $\theta_\mathrm{B}$, the harmonic quantum fluctuations may turn the saddle point into minima, whereby stabilizing the collinear states. This may occur when $\theta_\mathrm{B}\sim 1/S$, which is a vanishingly small window when $S\gg 1$. By contrast, all of the three dipolar magnetic ordered states discussed in Sec.~\ref{sec:obd_phases} occupy a finite interval of the $\theta_\mathrm{B}$ axis.

Given the rich physics displayed by this model, a natural question is  whether there exist materials that may potentially realize Eq.~\eqref{eq:model} as a first approximation. We first note that the pyrochlore lattice does not possess bond-inversion symmetry and hence the Dzyaloshinskii-Moriya (DM) interaction is always allowed. The DM interaction lifts the extensive degeneracy of the bilinear Heisenberg model~\cite{Elhajal2005,Cannals2008,Chern2010}. This observation suggests that one should focus on $3d^5$ (half-filled $d$ shell) transition metal ions where the spin-orbital interaction is typically the weakest~\cite{Khomskii}. Secondly, one must look for systems with a sizeable positive biquadratic interaction, $B$. While a negative biquadratic interaction often originates from magnetoelastic interactions~\cite{Kittel1960}, electronic correlations may produce a biquadratic interaction of either sign. In the latter case, both the sign and the magnitude of $B$ are sensitive to the details of the material electronic structure~\cite{Nagaev1982}. First-principle electronic structure calculations would be helpful in guiding the search for suitable candidate materials. Encouragingly, there exists a report~\cite{Sadeghi2015} based on first-principle calculations that FeF$_3$, where $S=5/2$ Fe$^{3+}$ ions reside on a pyrochlore lattice, possesses $B>0$ and a DM interaction slightly \emph{weaker} than $BS^2$. A very recent paper~\cite{Sanders_NaSrMn} 
reports the synthesis and characterization of the pyrochlore antiferromagnet NaSrMn$_2$F$_7$ with $S=5/2$ Mn$^{2+}$ ions. The Na/Sr disorder causes this material to enter the spin glass phase below a very low spin freezing temperature. It would be interesting to ascertain if it displays some aspects of the octupolar or dipolar correlations discussed in the present work above the freezing temperature. Finally, we note that there is also an alternative route toward positive biquadratic interaction through the double exchange mechanism. It is known that double-exchange generates a negative (ferromagnetic) bilinear coupling and a positive biquadratic coupling in Eq.~\eqref{eq:model}~\cite{Cieplak1978,Ishizuka2015a}. If one could find a material in which the ferromagnetic bilinear coupling resulting from double-exchange is much weaker than the antiferromagnetic bilinear coupling due to superexchange, then Eq.~\eqref{eq:model} could be regarded as a pertinent effective model Hamiltonian~\cite{Motome2010}.

To conclude, we believe our study has merely uncovered a small part of the many interesting properties of the  bilinear-biquadratic pyrochlore Heisenberg antiferromagnet model. We hope that our results instill theoretical interests in this model and motivate a search for candidate materials along with a systematic investigation of their properties.


\begin{acknowledgments}
We thank Vladislav Borisov, Gia-Wei Chern, Sasha Chernyshev, Hiroaki Ishizuka, Harald Jeschke, Yong Baek Kim, Roderich Moessner, Karlo Penc, and Roser Valent\'i for useful discussions. This research was supported in part by the Perimeter Institute for Theoretical Physics. Research at the Perimeter Institute is supported by the Government of Canada through Innovation, Science and Economic Development Canada and by the Province of Ontario through the Ministry of Research, Innovation and Science. The work at the U. of Waterloo was supported by the Canada Research Chair program (M.G., Tier 1).
\end{acknowledgments}


\appendix

\section{Counting weathervane modes}\label{app:counting}

\begin{figure}
\includegraphics[width=\columnwidth]{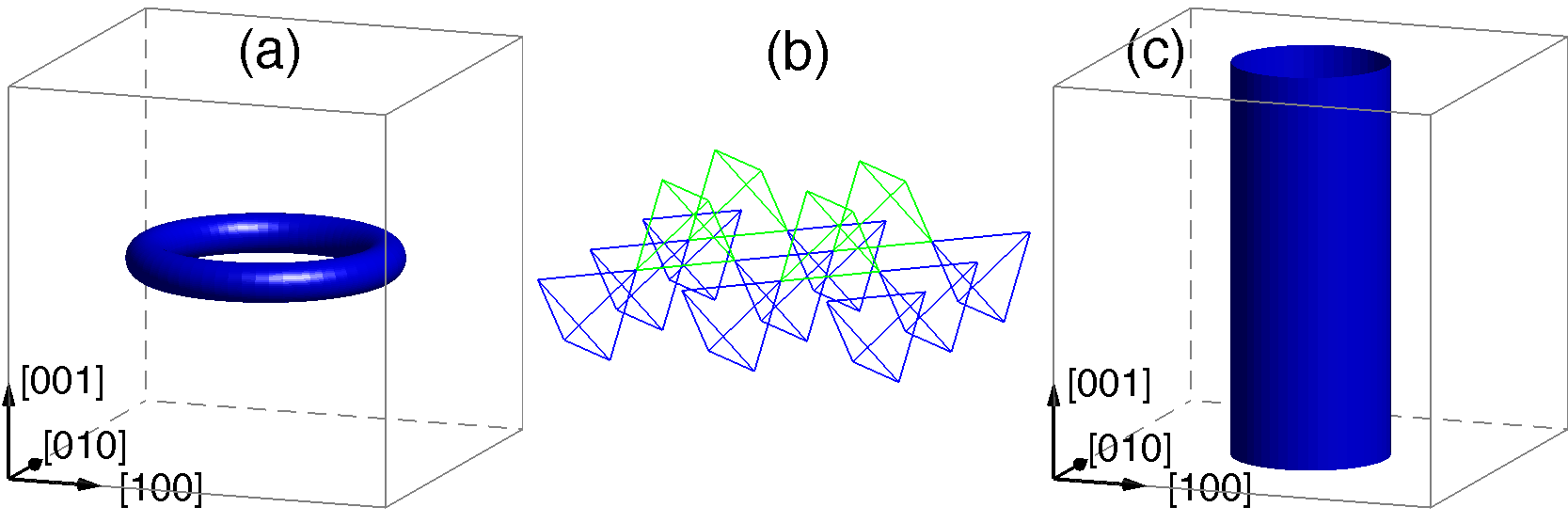}
\caption{(a) A system with open boundary conditions in all three directions. The blue torus presents a possible weathervane membrane localized in the system. (b) Two $(001)$ layers in the pyrochlore lattice. The bottom layer (blue) and the neighboring layer (green) are shown. (c) A system with open boundary condition in $[100]$ and $[010]$ and periodic boundary condition in $[001]$. The open boundary surfaces are $(100)$ and $(010)$. The blue cylinder represents a possible weathervane membrane that percolates along the crystallographic cubic $[001]$ direction.}
\label{fig:counting}
\end{figure}

In Sec.~\ref{sec:weathervane}, we showed that the color ice states support two-dimensional weathervane modes. An important question is how many such  modes a given color ice state supports. Below, we give a partial answer to this question. Firstly, we prove that the weathervane membrane, the two-dimensional  structure that supports a weathervane mode, must percolate through the system. As a corollary, the number of weathervane modes in a periodic system of linear dimension $L$ cannot be $O(L^3)$. Secondly, we argue that the counting of weathervane modes is unlikely to be $O(L^2)$. Finally, we provide numerical evidence for the claim that the counting of weathervane modes cannot exceed $O(L)$.

\emph{No $O(L^3)$ scaling}  --
To begin, we consider a pyrochlore lattice with open boundaries. Specifically, we take the open surfaces to be $(100)$, $(010)$, and $(001)$ surfaces (Fig.~\ref{fig:counting}a). We assume the spins are in a classical ground state. In other words, every tetrahedron is in TSS. We now prove by {\it reducio ad absurdum} that the weathervane membrane must percolate in this system. Suppose the opposite is true, and that there is a weathervane membrane localized inside the system. In such a case, if we fix the orientation of all spins on the six boundaries, the dimension ${\cal D}$ of the classical ground state manifold in such an open boundary system supporting a weathervane mode must be at least ${\cal D}=1$. However, such situation cannot occur. To see this, we start with the $(001)$ boundary layer at the bottom and suppose all the spins in this layer are fixed (Fig.~\ref{fig:counting}b, blue tetrahedra). We then consider the next $(001)$ layer (Fig.~\ref{fig:counting}b, green tetrahedra). On one hand, every tetrahedron in the second layer share two sites with the first layer. On the other hand, given that the four spins belonging to a tetrahedron are in TSS (Fig.~\ref{fig:tetra_ice}a), if two member spins are fixed, the other two spins cannot rotate continuously (but a reflection is allowed). Thus, the spins in the second layer cannot rotate continuously. By repeating this argument, we find that no spin can rotate continuously if we fix the boundary spins and, therefore, the classical ground state manifold dimension is ${\cal D}=0$, and not ${\cal D}\ge 1$ as demanded above.

We have therefore proven that a localized weathervane membrane does not exist and that, instead, it must percolate through the boundaries. An immediate corollary is that the number of weathervane modes \emph{cannot} be of order $L^3$ in a periodic lattice of linear dimension $L$. This follows from the observation that a number $O(L^3)$ of weathervane modes would require a localized weathervane membrane (of finite size) and whose spatial positioning in the system would then generate $O(L^3)$ weathervane modes. The latter situation reminds one of the weathervane mode about the so-called $\sqrt{3}\times\sqrt{3}$ ground state in the two-dimensional classical Heisenberg kagome antiferromagnet~\cite{Chalker1992}. In that case, those modes correspond to localized zero-energy excitations that reside on $6$-site hexagons, with their number being proportional to the number of the sites in the system.

\emph{No $O(L^2)$ scaling} -- Having ruled out an $O(L^3)$ scaling, one is next led to ask whether the counting of weathervane modes could be $O(L^2)$. The existence of $O(L^2)$ weathervane modes would require that the weathervane membrane percolates through the lattice in one direction while being localized in the other two directions (Fig.~\ref{fig:counting}c). We call the direction along which the weathervane membrane percolates the ``percolation direction''. We now show that the percolation direction cannot be $[001]$. By symmetry, the percolation direction cannot be $[100]$ and $[010]$ either. 

We consider a lattice with periodic boundary conditions along $[001]$ and open boundary condition along $[100]$ and $[010]$. Specifically, $(100)$ and $(001)$ are the open boundary surfaces for this system. Suppose there is a weathervane membrane percolating along the $[001]$ direction. We would then be able to contain the membrane within our system. As a result, the classical ground state manifold dimension would be $\mathcal{D}\ge 1$ after we fix the orientation of the spins on the open boundaries. However, by using the same line of reasoning as in the argument just above, we can show that the classical ground state manifold dimension is actually $\mathcal{D}=0$, which is a contradiction.

Likewise, we can rule out other high symmetry directions, namely $\langle 111\rangle$ and $\langle 110\rangle$, as percolation directions. We therefore argue that the counting of weathervane modes is unlikely to be $O(L^2)$.

\emph{Plausibility of $O(L^1)$ scaling} -- Finally, we performed a direct numerical enumeration of the weathervane modes. We consider color ice states commensurate with the $3\times 3\times 3$ rhombohedral supercell (Sec.~\ref{sec:obd_ice}) and stack them into a $6\times 6\times 6$ grid. We generate more than $3\times 10^6$ random color ice states by using loop updates~\cite{Chern2011,Khemani2012,Chern2014}, and enumerate the weathervane modes within each color ice state by using the procedure detailed in Sec.~\ref{sec:weathervane}. We find that the maximal number of weathervane modes is attained by the AIAO states. Since the AIAO states in a system of linear dimension $L$ has $O(L)$ weathervane modes, our result therefore strongly suggests the counting cannot exceed $O(L)$.


\section{A special property of the dynamical matrix}
\label{app:dyna_sym}

In Sec.~\ref{sec:obd_wv}, we stated that the dynamical matrices $\mathrm{M}$ in two classical ground states are identical up to a gauge transformation if these two ground states are related by a weathervane mode with rotation angle $\phi_\mathrm{wv}=\pi$. Here, we prove this statement and discuss its implications.

Our strategy is as follows. We consider two classical ground states, dubbed GS1 and GS2, and the associated dynamical matrices $\mathrm{M}_1$ and $\mathrm{M}_2$. GS2 is related to GS1 by a weathervane mode with $\phi_\mathrm{wv}=\pi$. We utilize the gauge covariance of $\mathrm{M}$: The explicit form of $\mathrm{M}$ depends on the local frame $\{\hat{x}_i,\hat{y}_i,\hat{z}_i\}$. While $\hat{z}_i$ is fixed by the spin orientation in a classical ground state, we are free to choose $\hat{x}_i,\hat{y}_i$, which amounts to a gauge choice since it would not affect any of the physical observables. Here, we shall choose the frames in GS2 such that $\mathrm{M}_1=\mathrm{M}_2$. Note that we ought to maintain the \emph{right-handedness} of the spin frames.

GS1 and GS2 differ by a $\phi_\mathrm{wv}=\pi$ weathervane mode that is localized on a weathervane membrane. Therefore, only the local frames on sites belonging to the weathervane membrane take different orientation between these two classical ground states. Consequently, $(\mathrm{M}_1)_{i\alpha,j\beta}$ and $(\mathrm{M}_2)_{i\alpha,j\beta}$ may be potentially different if and only if $i$ or $j$ is on the membrane.

We first consider the case in which $i$ is on the weathervane membrane while $j$ is off the membrane. Let $\{\hat{x}_i,\hat{y}_i,\hat{z}_i\}$ and $\{\hat{x}_j,\hat{y}_j,\hat{z}_j\}$ be the orthonormal frames at site $i$ and $j$ in GS1, respectively. In particular, $\hat{z}_i$ and $\hat{z}_j$ indicate the spin orientation at $i$ and $j$ in GS1. We recall that, in a weathervane mode, the spins on the weathervane membrane rotate with respect to a common axis. The axis coincides with the orientation of unrotated spins. Here, the spin $\hat{z}_i$ rotates with respect to $\hat{z}_j$ by $\pi$ in the weathervane mode. We thus obtain GS2 from GS1 by the said $\pi$ rotation. In particular, the frame on $i$ becomes,
\begin{subequations}
\begin{align}
\hat{x}'_i &= 2(\hat{x}_i\cdot\hat{z}_j)\hat{z}_j-\hat{x}_i,\nonumber\\
\hat{y}'_i &= 2(\hat{y}_i\cdot\hat{z}_j)\hat{z}_j-\hat{y}_i,\nonumber\\
\hat{z}'_i &= 2(\hat{z}_i\cdot\hat{z}_j)\hat{z}_j-\hat{z}_i,
\end{align}
after the rotation. We could use the above as the frame on site $i$ in GS2. However, we utilize the aforementioned gauge freedom and choose a different frame. Specifically, we rotate the $\hat{x}'$ and $\hat{y}'$ by $\pi$ with respect to $\hat{z}'_i$,
\begin{align}
\hat{x}''_i &= \hat{x}_i-2(\hat{x}_i\cdot\hat{z}_j)\hat{z}_j,\nonumber\\
\hat{y}''_i &= \hat{y}_i-2(\hat{y}_i\cdot\hat{z}_j)\hat{z}_j,\nonumber\\
\hat{z}''_i &= 2(\hat{z}_i\cdot\hat{z}_j)\hat{z}_j-\hat{z}_i.
\end{align}
We take $\{\hat{x}''_i,\hat{y}''_i,\hat{z}''_i\}$ to be the frame attached to $i$ in GS2. In particular, since all aforementioned rotations are \emph{proper rotations}, $\{\hat{x}''_i,\hat{y}''_i,\hat{z}''_i\}$ is a right-handed frame. Note the frame on $j$ in GS2 is the same as in GS1, namely $\{\hat{x}_j,\hat{y}_j,\hat{z}_j\}$, since the spin on site $j$ is unrotated.

Given the frames, we now compare $g_{i\alpha,j\beta}$, $g_{i\alpha,jz}$, and $g_{iz,j\beta}$ in GS1 and GS2. We find
\begin{align}
g_{i\alpha,j\beta} &= g''_{i\alpha,j\beta},\nonumber\\
g_{i\alpha,jz} &= -g''_{i\alpha,jz},\nonumber\\
g_{iz,j\alpha} &= -g''_{iz,j\alpha}.
\end{align}
Here, the left hand side is for GS1 whereas the right hand side is for GS2. Substituting the above into Eq.\eqref{eq:mentry}, we obtain
\begin{align}
(\mathrm{M}_1)_{i\alpha,j\beta} = (\mathrm{M}_2)_{i\alpha,j\beta}.
\end{align}
\end{subequations}

Likewise, we may consider the case in which both $i$ and $j$ are on the weathervane membrane and show that $(\mathrm{M}_1)_{i\alpha,j\beta}=(\mathrm{M}_2)_{i\alpha,j\beta}$. Putting both cases together, We therefore have $\mathrm{M}_1=\mathrm{M}_2$.

The above result implies that GS1 and GS2 have the same spin wave spectrum and the same zero point energy within the quadratic approximation (Eq.~\eqref{eq:zpe}). We remark that the property of the dynamical matrix $\mathrm{M}$ discussed here is reminiscent of the situation in the classical kagome Heisenberg antiferromagnet, where the dynamical matrix is identical for all coplanar states~\cite{Chalker1992}


\section{Zero modes and soft modes}\label{app:softmode}

\begin{figure}
\includegraphics[width=\columnwidth]{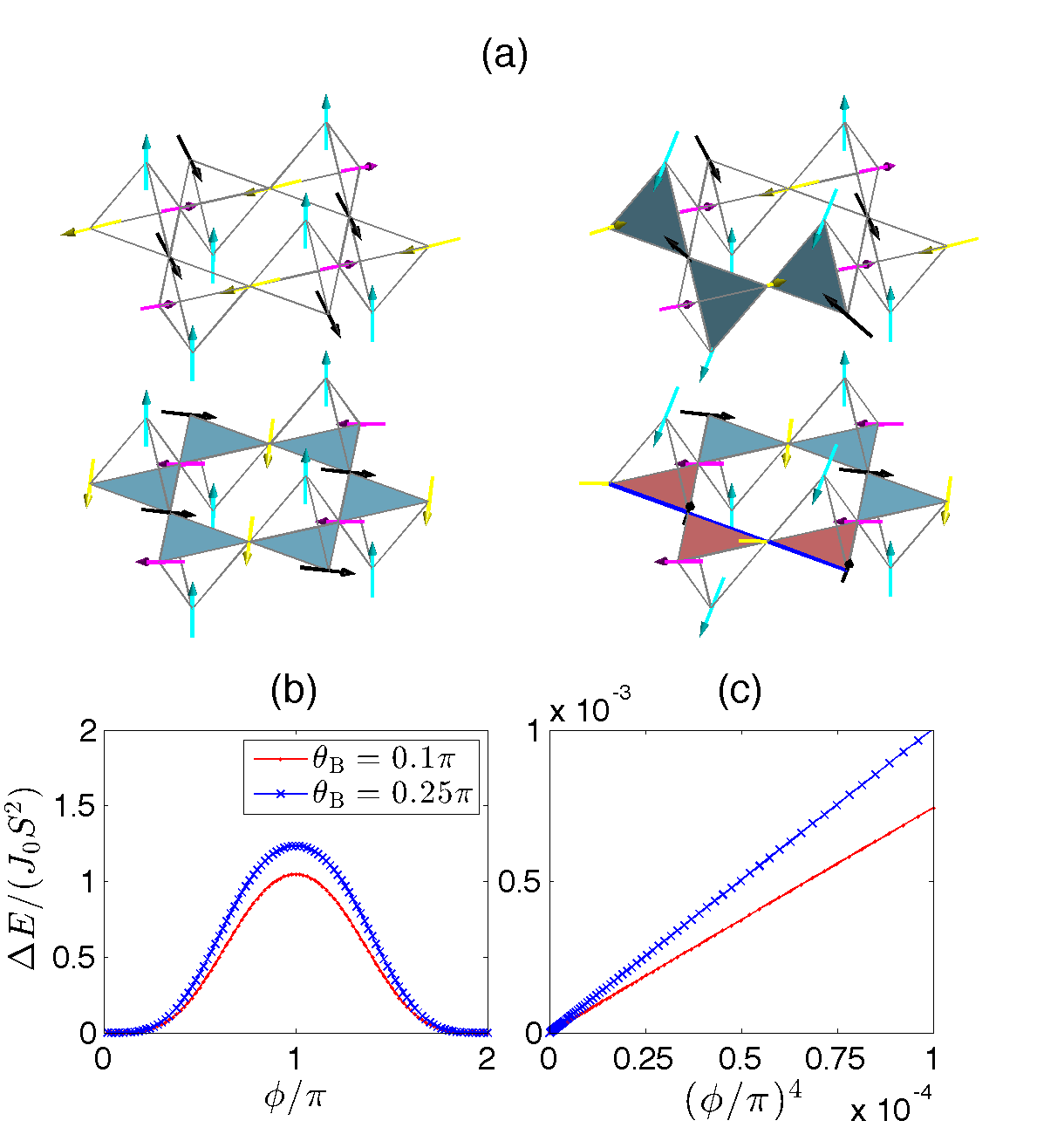}
\caption{(a) Top left: an AIAO state. Spins colored in cyan, magenta, yellow, and black point along $\mathbf{n}_\mathrm{C}$, $\mathbf{n}_\mathrm{M}$, $\mathbf{n}_\mathrm{Y}$, and $\mathbf{n}_\mathrm{K}$, respectively. Top right: A color-companion (CC) of the AIAO state. Shaded triangles highlight the weathervane membrane. Spins belonging to the weathervane membrane are rotated by an angle $\pi$ with respect to $\mathbf{n}_\mathrm{M}$. Bottom left: a weathervane mode supported by the AIAO state. Spins on the shaded traingles may collectively rotate with respect to $\mathbf{n}_\mathrm{C}$ by an arbitrary angle $\phi$ at no energy cost. We set $\phi=\pi/2$ for the visulization purpose. Bottom right: A soft mode supported by the CC state. Spins covered by the shaded triangles rotate by an angle $\phi$ with respect to various axes (see the main text for details). Here, we arbitrarily set $\phi=\pi/2$. Tetradra with red bottom are not in a tetrahedral spin state (TSS). (b) Classical energy cost $\Delta{}E$ per defect tetrahdron (i.e., a tetrahedron that is no longer in a TSS) as a function of rotation angle $\phi$. We set $\theta_\mathrm{B}=0.1\pi$ (red dots) and $\theta_\mathrm{B}=0.25\pi$ (blue crosses). (c) Behavior of $\Delta{}E$ as a function of $\phi$ in the limit of $\phi\to 0$.}
\label{fig:soft_mode}
\end{figure}

In Sec.~\ref{sec:weathervane}, we showed that a color ice state may support weathervane modes, which are genuine zero-energy collective spin rotations. In Sec.~\ref{sec:spinwave}, we pointed out that weathervane modes imply the dynamical matrix $\mathrm{M}$ (Sec.~\ref{sec:obd_formalism}) has zero eigenvalues. However, the converse is not true; a classical ground state with zero eigenvalues in its $\mathrm{M}$ does not necessarily support true zero-energy excitations beyond quadratic order. Exciting a zero eigenmode of $\mathrm{M}$ costs no classical energy within the quadratic approximation. Once we go beyond the quadratic approximation, the energy cost for exciting such a mode may be nonzero. It is therefore necessary to classify the zero eigenmodes of $\mathrm{M}$ into two categories. To borrow the terminology used for the classical kagome Heisenberg antiferromagnet~\cite{Chalker1992}, we refer to the zero eigenmodes of $\mathrm{M}$ which correspond to genuine zero-energy collective spin rotations as \emph{zero-modes} and those that do not as \emph{soft modes}. As we shall see below, the classical energy cost for exciting a soft mode is proportional to the fourth power of its amplitude in the small amplitude limit.

The weathervane modes discussed in Sec.~\ref{sec:weathervane} are zero modes by definition and need no further elaboration. The best way to illustrate the soft modes is through a concrete example. To this end, we begin with an AIAO state (Fig.~\ref{fig:soft_mode}a, top left panel). In Sec.\ref{sec:weathervane}, we showed that an AIAO state supports an infinite number of weathervane modes. Each weathervane mode is localized onto a kagome layer. We pick a given kagome layer and then rotate all the spins in this layer with respect to $\mathbf{n}_\mathrm{M}$ by an angle $\pi$ (Fig.~\ref{fig:soft_mode}a, top right panel). The resulting state is a new ground state; a color-companion (CC) of the AIAO state (Sec.~\ref{sec:obd_wv}). Hereafter, we refer to this state simply as the CC state.

We now show that the CC state supports soft modes. Our strategy is as follows. We utilize the fact that the dynamical matrix of the CC state in appropriately chosen spin frames is identical to that of the AIAO state (App.~\ref{app:dyna_sym}). Therefore, a zero eigenmode of the former dynamical matrix corresponds to a zero eigenmode of the latter. We then extend the zero eigenmode supported by the CC state, which are infinitesimal spin rotations, to finite rotations and demonstrate that it costs nonzero energy.

We consider a weathervane mode in the AIAO state. In the example shown in the bottom left panel of Fig.~\ref{fig:soft_mode}a, the spins belong to a given kagome layer (shaded triangles) rotated collectively with respect to $\mathrm{n}_\mathrm{C}$ by the same angle $\phi$. Let $\{\hat{x}_i,\hat{y}_i,\hat{z}_i\}$ be the spin frame attached to a kagome site $i$ in the AIAO state. In the limit of $\phi\to 0$, the resulting spin orientation is given by $\mathbf{n}_i=x_i\hat{x}_i+y_i\hat{y}_i+\hat{z}_i$, where,
\begin{align}
x_i = \phi(\mathbf{n}_C\times\hat{z}_i)\cdot\hat{x}_i,\quad y_i = \phi(\mathbf{n}_C\times\hat{z}_i)\cdot\hat{y}_i.
\label{eq:xylist}
\end{align}
The list of $\{x_i,y_i\}$ for all sites $i$ belonging to the kagome layer describes a zero eigenmode of the dynamical matrix for the AIAO state.

To find the corresponding zero eigenmode of the dynamical matrix for the CC state, we use the results in Appendix~\ref{app:dyna_sym}. Recall the CC state is obtained from the AIAO state by a $\pi$ rotation with respect to $\mathrm{n}_\mathrm{M}$ on a subset of spins (Fig.~\ref{fig:soft_mode}a, top right panel). To ensure that the dynamical matrix for the CC state and the AIAO states is the same, we need to choose the spin frames in the CC state properly. Specifically, if a spin on a site $i$ is rotated, we set the frames in the CC state to be $\hat{x}'_i=-\mathrm{R}(\mathbf{n}_\mathrm{M},\pi)\hat{x}_i$, $\hat{y}'_i=-\mathrm{R}(\mathbf{n}_\mathrm{M},\pi)\hat{y}_i$, and $\hat{z}'_i=\mathrm{R}(\mathbf{n}_\mathrm{M},\pi)\hat{z}_i$. Here, $\mathrm{R}(\mathbf{n}_\mathrm{M},\pi)$ stands for the rotation matrix with respect to $\mathbf{n}_\mathrm{M}$ by angle $\pi$, and $\{\hat{x}_i,\hat{y}_i,\hat{z}_i\}$ is the spin frame in the AIAO state. On the other hand, if the spin on site $i$ is not rotated, we set the spin frame attached to $i$ in the CC state to be the same as in the AIAO state, i.e. $\{\hat{x}_i,\hat{y}_i,\hat{z}_i\}$. 

With this properly chosen system of spin frames, the dynamical matrix of the CC state is identical to that of the AIAO state. Hence, the previously prescribed list of $\{x_i,y_i\}$, where $i$ runs over all sites of the \emph{same} kagome layer (Fig.~\ref{fig:soft_mode}a, bottom right panel), describe a zero eigenmode of the dynamical matrix for the CC state. 

To determine the spin orientation in the zero eigenmode, we use the previously chosen spin frames. In particular, if the kagome site $i$ was previously rotated during the construction process of the CC state from the AIAO state (sites covered by the thick blue line in the bottom right panel of Fig.~\ref{fig:soft_mode}a), the spin orientation is given by
\begin{align}
\mathbf{n}_i & = x_i\hat{x}'_i+y_i\hat{y}'_i+\hat{z}'_i\nonumber\\
& = -x_iR(\mathbf{n}_\mathrm{M},\pi)\hat{x}_i-y_iR(\mathbf{n}_\mathrm{M},\pi)\hat{y}_i+R(\mathbf{n}_\mathrm{M},\pi)\hat{z}_i\nonumber\\
& = \mathrm{R}(\mathbf{n}_\mathrm{M},\pi)[\hat{z}_i-\phi(\mathbf{n}_C\times\hat{z}_i)]\nonumber\\
& \approx \mathrm{R}(\mathbf{n}_\mathrm{M},\pi)\mathrm{R}(\mathbf{n}_\mathrm{C},-\phi)\hat{z}_i.
\end{align}
In the third line, we plug Eq.~\eqref{eq:xylist} in and recognize that the operation in the square bracket is an infinitesimal rotation with respect to $\mathbf{n}_\mathrm{C}$. In the last line, we have upgraded it to a finite rotation. Otherwise, if a kagome site $i$ does not get rotated during the AIAO$\rightarrow$CC construction, the spin orientation is given by
\begin{align}
\mathbf{n}_i &= x_i\hat{x}_i+y_i\hat{y}_i+\hat{z}_i\nonumber\\
& = \hat{z}_i+\phi(\mathbf{n}_C\times\hat{z}_i)\approx \mathrm{R}(\mathbf{n}_\mathrm{C},\phi)\hat{z}_i.
\end{align}
We thus have explicitly constructed a zero eigenmode of the dynamical matrix for the CC state.

We now demonstrate the aforementioned eigenmode indeed is a soft mode. We first show that it costs finite classical energy when $\phi$ is not small. Inspection of the resulting state (Fig.~\ref{fig:soft_mode}a, bottom right panel) shows that performing the aforementioned rotation on the CC state brings some of the tetrahedra out of a TSS (the tetrahedra with red bottom). Therefore, the classical energy increases. This is confirmed by a direct evaluation of the classical energy cost $\Delta E(\phi)$ (Fig.~\ref{fig:soft_mode}b). We show in Fig.~\ref{fig:soft_mode}c, $\Delta E(\phi) \propto \phi^4$ when $\phi$ is small. This confirms our previous claim that the energy cost for exciting a soft mode is a quartic function of its amplitude when the amplitude is small.

\bibliography{pyro_biquadratic.bib}

\end{document}